\documentclass[11pt,preprint]{aastex} 
\usepackage{amsmath,amssymb,amsfonts} 
\usepackage{color}                    
\usepackage{epsfig}
\usepackage{amsbsy} 
\newcommand{\bu}{\boldsymbol{u}}

\newcommand{\bomega}{\boldsymbol{\omega}}
\newcommand{\bdot}{\boldsymbol{\cdot}}
\newcommand{\bnabla}{\boldsymbol{\nabla}}

\def\t{\theta}
\def\b{\beta}
\def\cs{c_{\rm s}}

\slugcomment{to appear in MNRAS (accepted June 16, 2011)}

\shorttitle{Dynamical Friction in a Gas}
\shortauthors{Lee \& Stahler}

\begin{document}

\title{Dynamical Friction in a Gas: The Subsonic Case}

\author{Aaron T. Lee\altaffilmark{1} and Steven W. Stahler\altaffilmark{1}}

\altaffiltext{1}{Astronomy Department, University of California,
Berkeley, CA 94720}

\email{a.t.lee@berkeley.edu}

\begin{abstract}
We study the force of dynamical friction acting on a gravitating point mass that 
travels through an extended, isothermal gas. This force is well established 
in the hypersonic limit, but remains less understood in the subsonic regime. 
Using perturbation theory, we analyze the changes in gas velocity and density 
far from the mass. We show analytically that the steady-state friction force is 
${\dot M}\,V$, where $\dot M$ is the mass accretion rate onto an object moving
at speed $V$. It follows that the speed of an object experiencing no other forces 
declines as the inverse square of its mass. Using a modified version of the classic Bondi-Hoyle interpolation formula for $\dot M$ as a function of $V$, we derive an analytic expression for the 
friction force. This expression also holds when mass accretion is thwarted, e.g. by a wind, as long as the wind-cloud interaction is sufficiently confined spatially. Our result should find application in a number of astrophysical settings, such as the motion of galaxies through intracluster gas.
\end{abstract}


\section{Introduction}
\label{sec:intro}

A gravitating mass that traverses a sea of other particles builds up an 
overdense wake behind it. This wake tugs back on the mass, providing an 
effective drag. The background sea may itself consist of non-interacting point 
masses. For this collisionless case, \citet{c43} first derived the dynamical 
friction force. His celebrated result has since found application in a great 
many astrophysical problems, ranging from mass segregation in dense star 
clusters \citep{pm02} to planet migration through interaction with 
planetesimals \citep{dye03}.

The background environment may also be an extended gas cloud. This type of 
dynamical friction has also been invoked in a variety of contexts, such as 
black hole mergers in galactic nuclei \citep{d06}, the heating of the 
intracluster medium by infalling galaxies \citep{e04}, and the migration of giant planets within a protoplanetary disk \citep{odi10}. When the ambient medium
is a gas, pressure gradients influence the formation of the wake behind the 
gravitating object. Surprisingly, the general determination of  
gaseous dynamical friction, for arbitrary Mach number of the perturbing mass, 
has not yet been achieved. There is substantial agreement when the mass is 
traveling hypersonically with respect to the gas. In this limit, the force 
varies as $V^{-2}$, where $V$ is the speed of the perturber 
\citep{d64,rs71,rs80,o99}.   

In all studies thus far, the authors first calculated the properties of the  
wake by treating it as a linear perturbation of the background gas. The density perturbation is symmetric upstream and downstream when the mass is moving subsonically \citep{d64}, leading \citet{rs80} to conclude that the friction force is zero in this case. \citet{o99} obtained the force through direct 
integration over surrounding fluid elements, using their respective density 
enhancements. She added the constraint that the projectile's gravitational 
field only be switched on for a finite time interval $\Delta t$. With this 
device, she first found a nonzero result even in the subsonic regime. The force increases with $V$, and logarithmically diverges at a Mach number of unity.

Interestingly, the quantity $\Delta t$ does not appear in Ostriker's final expression for the subsonic force. This fact indicates that the artifice of a finite time interval 
was unnecessary and that a steady-state analysis is applicable. Indeed, the
force attains a steady-state value in the numerical simulations of 
\citet{sb99}. The divergence at a Mach number of unity in the
analytical expression further suggests that physical understanding of the
problem is incomplete.

In this paper, we revisit the subject of dynamical friction, concentrating
entirely on the less studied subsonic case. We take the perturbing body to be
a point mass $M$ traveling through an initially uniform gas. The previous 
studies cited also ostensibly dealt with point masses, in the sense that the 
physical size of the body was ignored. However, it was assumed, either
tacitly or explicitly, that the object's radius $R$ far exceeds the accretion 
radius $r_{\rm acc}$, conventionally defined as 
\hbox{$r_{\rm acc}\,\equiv\,2\,G\,M/V^2$}. It is true that when 
\hbox{$R \,\gg\,r_{\rm acc}$}, the gravitational force from the object is so 
weak that mass accretion by infall is negligible. Under these circumstances, however,
the primary drag on the body is not from dynamical friction, but from direct
impact by the gas, a fact sometimes overlooked.\footnote{\citet{rs71} recognized that both drag forces act on galaxies moving supersonically through intracluster gas (see their eq. 5). They extended the linear analysis of the flow into the nonlinear regime, utilizing a similarity solution. However, their focus was the X-ray emission from the wake and bowshock, rather than the actual motion of the galaxies.} 

The conventionally assumed inequality marginally holds in one situation 
commonly envisioned, galaxies within intracluster gas 
(\hbox{$R\,\sim\,r_{\rm acc}\,\sim\,10^{24}\,\,{\rm cm}$}). However, it fails badly in other contexts, e.g., supermassive black holes within galaxies 
(\hbox{$R\,\sim\,10^{11}\,\,{\rm cm}$}, \hbox{$r_{\rm acc}\,\sim\,10^{19}\,\,{\rm cm}$}) 
or gas giant planets inside circumstellar disks (\hbox{$R\,\sim\,10^{10}\,\,{\rm cm}$},
\hbox{$r_{\rm acc}\,\sim\,10^{13}\,\,{\rm cm}$}). When \hbox{$R\,\ll\,r_{\rm acc}$}, 
as we assume here, dynamical friction is indeed the main drag force. The 
relative density enhancement in the wake is not small, as needed for linear 
theory \citep[see, e.g.,][]{kk09}, and mass accretion cannot be neglected.

Our analysis indeed pivots on the fact that the transfer of linear momentum
from the background gas to the object, which underlies the friction force, is 
closely related to the transfer of mass. The problem of gas accretion onto a 
moving body was addressed in a classic series of papers by \citet{hl39},
\citet{bh44}, and \citet{b52}. The final result for the accretion rate, 
applicable for all Mach numbers, is the interpolation formula offered by 
\citet{b52}. While not derived rigorously, the formula matches known results 
in the hypersonic and stationary limits, and is broadly consistent with numerical simulations \citep[see][and references therein]{r96}.

The strategy in our paper is to determine, using perturbation theory, the
density and velocity of the gas. However, we focus not on the wake, as in previous studies, but on 
a region {\it far} from the object, where its gravity is relatively weak.
Extending the perturbation analysis into the nonlinear regime, we calculate the net momentum flux onto the accreting object and derive
analytically that the force from dynamical friction is $\dot M\,V$, where 
$\dot M$ is the mass accretion rate onto the object. Adopting an analytic form for this rate, the drag force follows. This force first rises with $V$ and then falls, remaining finite at all Mach numbers. Moreover, there is 
a contribution from the direct accretion of fluid momentum onto the body. This 
contribution is absent in the stellar dynamical problem, but is here comparable
to the gravitational tug from the wake. 

In Section~\ref{sec:method} below, we introduce a perturbative series expansion to analyze the
far-field density and velocity. In Section~\ref{sec:outer}, we use this expansion to derive 
a hierarchy of dynamical equations, of which we need only solve the first two 
sets. Section~\ref{sec:mass} shows how the mass accretion rate is connected to solutions of 
our second-order equations. In Section~\ref{sec:friction}, we similarly relate the friction 
force to these solutions, and derive the central connection between this force 
and the accretion rate. Using a modified version of the Bondi interpolation formula for the latter, we find 
explicitly the deceleration of an isolated mass in Section~\ref{sec:velocity}. Finally, Section~\ref{sec:summary} 
compares our result with existing numerical simulations and suggests future 
avenues of inquiry.

\section{Outer Flow: Method of Solution}
\label{sec:method}
\subsection{Physical Assumptions}

\begin{figure}
\plotone{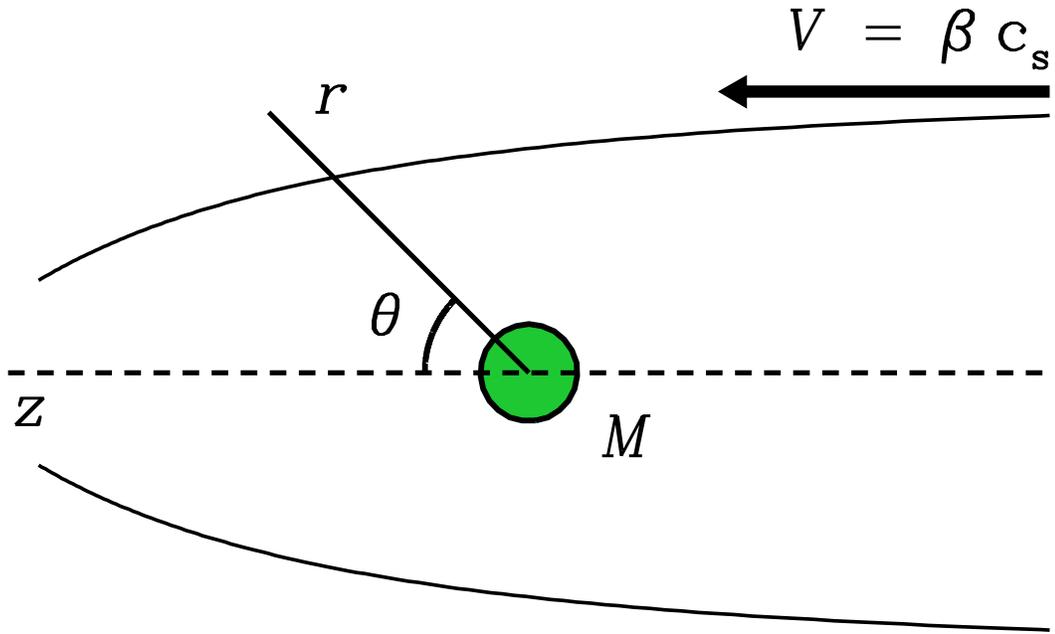}
\caption{Spherical coordinate system centered on a gravitating body of mass $M$. The gas is isothermal, and its velocity far upstream is $\b\,\cs\ (\b<1)$. The upstream direction corresponds to $\t=\pi$, and downstream to $\t=0$. While the figure shows the mass to have a finite physical size, we assume it to be a point particle in our analysis.}
\label{fig:coord}
\end{figure}

Let the gravitating mass $M$ travel in a straight line with speed $V$ through the 
extended gas cloud. Following previous analytic studies of dynamical friction, we assume the gas to be isothermal, with an associated sound speed 
$c_s$. Very far from the mass, the density is spatially uniform and has
the value $\rho_0$. We choose a reference frame whose origin is anchored on the
perturbing mass. In this frame, it is the gas that has speed $V$ far from 
the mass. We let the gas velocity be directed along the $z$-axis, and employ 
spherical coordinates $r$ and $\theta$ (see Fig.~\ref{fig:coord}). We now seek a steady-state, axisymmetric solution for the flow, which is
taken to be inviscid. We neglect the self-gravity of the gas, and assume that
each fluid element feels only a pressure gradient and the gravitational pull of the point mass. 

Strictly speaking, there is no steady flow, as this mass decelerates and $V$ 
continually changes \citep[e.g.,][]{f10}. What, then, is the meaning of the force we are calculating? Imagine the object being dragged by a massless string through the gas at the
fixed speed $V$. After a long time, a steady-state flow is indeed established 
throughout the surrounding gas, and the tension in the string approaches a 
constant value. This limiting tension is the dynamical friction force being
calculated here.

Return now to the actual case, in which there is no string and the mass
decelerates. As stated previously, there is no global, steady-state flow. The 
flow is quasi-steady, however, within some distance over which the altered 
motion of the mass is communicated by sound waves. We shall quantify this 
distance later, after we have established the flow assuming steady-state 
conditions.

One important property of the flow is that it is irrotational. Euler's 
equation in steady state may be written
\begin{equation}
\bu \times \bomega \,=\, \bnabla B  \,\,,
\end{equation}
where $\bu$ is the fluid velocity, \hbox{$\bomega \,\equiv\bnabla\times\bu$} is the vorticity, and the 
Bernoulli function $B$ is 
\begin{equation}
B \,\equiv\,{1\over 2}\,u^2 \,+\, 
c_s^2\,\,{\rm ln}\left(\,{\rho\over{\rho_0}}\right) \,-\,\frac{G\,M}{r}
\,\,.
\end{equation}
Both the fluid speed and density approach constant values far from the mass. Hence, $B$ is a spatial constant throughout the flow, and 
\begin{equation}
\bu \times \bomega \,=\, 0  \,\,.
\end{equation} 
Since $\bu$ is a poloidal vector, the vorticity $\bomega$ is toroidal. The last
equation then implies that \hbox{$\bomega\,=\,0$}, as claimed. We will not need
to invoke the irrotational character of the flow until Section~\ref{sec:friction}, when we explicitly
evaluate the dynamical friction force.

Throughout our analysis, it will be more convenient to employ, not the vector 
fluid velocity $\bu (r,\theta)$, but the scalar stream function 
$\psi (r,\theta)$. We may recover the individual velocity components from the 
stream function through the standard relations
\begin{eqnarray}
\label{eqn:ur}
u_r \,&=&\, {1\over{\rho\,r^2\,{\rm sin}\,\theta}}\,\,
{{\partial\psi}\over{\partial\theta}}\,\,, \\
\label{eqn:ut}
u_\theta\,&=&\,{-1\over{\rho\,r\,{\rm sin}\,\theta}}\,\,
{{\partial\psi}\over{\partial r}} \,\,,
\end{eqnarray}
where \hbox{$\rho\,=\,\rho (r,\theta)$} is the mass density. The velocity, as
given by equations~(\ref{eqn:ur}) and (\ref{eqn:ut}), automatically obeys mass continuity:
\begin{eqnarray}
0\,&=&\,\bnabla \bdot (\rho\,\bu)\,\,, \nonumber \\
0\,&=&\,{1\over r^2}\,{\partial{\phantom r}\over{\partial r}}
\left(\rho\,r^2\,u_r\right) \,+\, {1\over{r\,{\rm sin}\,\theta}}\,
{\partial{\phantom\theta}\over{\partial\theta}}
\left(\rho\,{\rm sin}\,\theta\,u_\theta\right) \,\,.
\end{eqnarray}

\subsection{Perturbation Expansion}
Far from the mass, as the density approaches $\rho_0$, the velocity has only a
$z$-component, which is $V$. Equivalently, we have in this limit 
\hbox{$u_r\,\approx\,V\,{\rm cos}\,\theta$} and
\hbox{$u_\theta\,\approx\,-V\,{\rm sin}\,\theta$}. It follows that the 
far-field limit of the stream function is
\begin{equation}\label{eqn:sfunc1}
\psi \,\approx \,{{\rho_0\,V\,r^2\,{\rm sin}^2\,\theta}\over 2} \,\,.
\end{equation}
For a more complete analysis of the flow in this region, we take equation~(\ref{eqn:sfunc1}) 
to represent the leading term of a perturbation expansion. Introducing the 
sonic radius \hbox{$r_s\,\equiv\,G\,M/c_s^2$}, we first rewrite equation~(\ref{eqn:sfunc1}) 
as 
\begin{equation}
\psi \,\approx\, \rho_0\,c_s\,r_s^2\,\beta\,
\left({r\over r_s}\right)^2\,{{{\rm sin}^2\,\theta}\over 2} \,\,,
\end{equation}
where $\beta$ is the Mach number of the projectile mass:
\begin{equation}
\beta \,\equiv\, {V\over c_s} \,\,.
\end{equation}

Our perturbation expansion of $\psi$ is then given by
\begin{equation}
\psi\,=\,\rho_0\,c_s\,r_s^2\,\left[
f_2\left(r\over r_s\right)^2 \,+\,
f_1\left(r\over r_s\right) \,+\,f_0\,+\,
f_{-1}\left(r\over r_s\right)^{-1} \,+\, ...\,\right] \,\,,
\end{equation}
where 
\begin{equation}\label{eqn:f2}
f_2 \,\equiv\,{{\beta\,\,{\rm sin}^2\,\theta}\over 2} \,\,, 
\end{equation}
and where $f_1$, $f_0$, $f_{-1}$, etc. are still unknown, nondimensional 
functions of $\beta$ and $\theta$. Similarly, we expand the density as
\begin{equation}
\rho \,=\, \rho_0\,\left[ 1\,+\,
g_{-1}\left({r\over r_s}\right)^{-1} \,+\,
g_{-2}\left({r\over r_s}\right)^{-2} \,+\,
g_{-3}\left({r\over r_s}\right)^{-3} \,+\,...\,\right] \,\,.
\end{equation}
Here, $g_{-1}$, $g_{-2}$, $g_{-3}$, etc. are also nondimensional functions of 
$\beta$ and $\theta$, all yet to be found. Both expansions are only valid for 
\hbox{$r\,\gg\,r_s$}, the inequality that defines our outer region. We further
assume that the physical radius of the object obeys \hbox{$R\,\ll\,r_s$}. Since
the motion is subsonic (\hbox{$V\,<\,c_s$}), it follows that 
\hbox{$R\,\ll\,r_{\rm acc}$}, so that mass accretion is significant.

At this point, it is convenient to cast all our variables into nondimensional 
form. We let the fiducial radius, density, and speed be $r_s$, 
$\rho_0$, and $\cs$, respectively. Similarly, the stream function is normalized to $\rho_0\,c_s\,r_s^2$. Then equations~(\ref{eqn:ur}) and
(\ref{eqn:ut}) relating the velocity to the stream function remain the same 
nondimensionally. We shall not employ  a new notation for nondimensional 
variables, but make it clear whenever we switch back to dimensional relations. 
With this convention, the nondimensional expansions for the stream function and
density simplify to
\begin{eqnarray}\label{eqn:psind}
\psi \,&=&\, f_2\ r^2\,\,+\,\,f_1\ r \,\,+\,\, f_0 \,\,+\,\, 
f_{-1}\ r^{-1}\,\,+\,... \,\,, \\ 
\label{eqn:rhond}
\rho \,&=&\,  1 \,\,+\,\,g_{-1}\ r^{-1} \,\,+\,\, g_{-2}\ r^{-2}\,\,+\,\,
g_{-3}\ r^{-3}\,\,+\,\,... \,\,.
\end{eqnarray}

By adopting these perturbation expansions, we have effectively limited our
analysis to the subsonic regime. For \hbox{$\beta\,>\,1$}, we expect the
fluid variables or their derivatives to be discontinuous across the Mach
cone, whose opening angle is given by 
\hbox{${\rm sin}\,\theta\,=\,\beta^{-1}$} \citep[see, e.g.,][]{rs71}. It would
thus be necessary to adopt two separate expansions for $\psi$ and $\rho$, 
one inside and one outside the Mach cone. To avoid this complication, and since
we are primarily interested in the subsonic regime in any event, we assume that
\hbox{$\beta\,<\,1$} and retain the single expansions.

\subsection{Boundary Conditions}
By symmetry, the upstream axis of the flow, defined by \hbox{$\theta\,=\,\pi$},
is a streamline for any $\beta$-value. That is, $\psi(r,\pi)$ is independent
of $r$. The actual value of $\psi (r,\pi)$ is immaterial, reflecting the fact
that the full function $\psi (r,\theta)$ can have an arbitrary additive 
constant without affecting the velocities. For convenience, we set
\hbox{$\psi (r,\pi)\,=\,0$}, and note from equation~(\ref{eqn:f2}) that $f_2 (\pi)$ 
already vanishes. From equation~(\ref{eqn:psind}) for the general expansion, we require that
\hbox{$f_i (\pi) \,=\,0$}, for \hbox{$i\,=\,1,\,0\,,-1,\,-2$}, etc.

A second set of boundary conditions pertains to the behavior of the velocity
$\bu$. Let us focus again on the upstream axis. The righthand
sides of both equations~(\ref{eqn:ur}) and (\ref{eqn:ut}) contain \hbox{${\rm sin}\,\theta$} in the
denominator. Since the density $\rho$ is finite at \hbox{$\theta\,=\,\pi$},
where \hbox{${\rm sin}\,\theta$} vanishes, both 
\hbox{$\partial\psi/\partial\theta$} and \hbox{$\partial\psi/\partial r$} must
tend to zero as $\theta$ approaches $\pi$, at least as fast as 
\hbox{${\rm sin}\,\theta$}.

Considering first \hbox{$\partial\psi/\partial\theta$}, we see that
\hbox{$\partial f_2/\partial\theta\,=
\,\beta\,{\rm sin}\,\theta\,{\rm cos}\,\theta$}, which properly vanishes. We 
must further demand that \hbox{$f_i^\prime\,(\pi)\,=\,0$}, for  
\hbox{$i\,=\,1,\,0\,,-1,\,-2$}, etc. Turning to 
\hbox{$\partial\psi/\partial r$}, the term involving $f_2$ still goes to zero, 
while \hbox{$f_0 (\theta)$} itself vanishes when taking the $r$-derivative of 
$\psi$. We are already requiring that \hbox{$f_i\,(\pi)\,=\,0$} for all other 
$i$. Thus, the stipulation that \hbox{$\psi\,(r, \pi)\,=\,0$} implies that  
\hbox{$\partial\psi/\partial r$} also vanishes at \hbox{$\theta\,=\,\pi$}, so
that $u_\theta$ does not diverge.

Approaching the downstream axis, \hbox{${\rm sin}\,\theta$} again vanishes as
$\theta$ goes to zero. By analogous reasoning, we require that
\hbox{$f_i^\prime\,(0)\,=\,0$} for \hbox{$i\,=\,1,\,0\,,-1,\,-2$}, etc. To
ensure the regularity of $u_\theta$, we further need \hbox{$f_i\,(0)\,=\,0$},
for \hbox{$i\,=\,1,\,-1,\,-2$}, etc. Again, the term $f_0$ disappears when
taking the $r$-derivative of $\psi$, and there is no a priori restriction on 
\hbox{$f_0\,(0)$}. Indeed, this quantity sets the mass accretion rate onto the
moving body, as we later demonstrate. In summary, our boundary conditions are:
\hbox{$f_i\,(\pi)\,=\,f_i^\prime\,(\pi)\,=\,f_i^\prime\,(0)\,=\,0$}, for 
\hbox{$i\,=\,1,\,0\,,-1,\,-2$}, etc., and \hbox{$f_i\,(0)\,=\,0$} for
\hbox{$i\,=\,1,\,-1,\,-2$}, etc.

\section{Outer Flow: Results}
\label{sec:outer}
\subsection{First-Order Equations}
Our inviscid flow obeys Euler's equation, which we write in spherical
coordinates. The $r$- and $\theta$-components of this vector equation are
\begin{eqnarray}\label{eqn:eulerr}
u_r\,{{\partial u_r}\over{\partial r}}\,+\,{u_\theta\over r}\,
{{\partial u_r}\over{\partial\theta}}\,-\,{{u_\theta^2}\over r}\,&=&\,
-{1\over\rho}\,{{\partial\rho}\over{\partial r}}\,-\,{1\over r^2}\,\,, \\
\label{eqn:eulert}
u_r\,{{\partial u_\theta}\over{\partial r}}\,+\,{u_\theta\over r}\,
{{\partial u_\theta}\over{\partial\theta}}\,+\,{{u_r\,u_\theta}\over r}\,&=&\,
-{1\over{\rho\,r}}\,{{\partial\rho}\over{\partial\theta}}\,\,,
\end{eqnarray}
where $u_r$ and $u_\theta$ are given in terms of $\psi$ by equations~(\ref{eqn:ur}) and
(\ref{eqn:ut}). Our strategy is to substitute the perturbation expansions~(\ref{eqn:psind}) and (\ref{eqn:rhond})
into Euler's equations. For each power of $r$, we demand that its coefficients
match. In this way, we will obtain a hierarchy of coupled equations for the
functions $f$ and $g$.  

Before proceeding, we first note that $u_r$ and $u_\theta$ are both
proportional to $\rho^{-1}$. To avoid expanding inverse powers of the density,
we multiply equations~(\ref{eqn:eulerr}) and (\ref{eqn:eulert}) through by $\rho^3$. Replacing the 
velocity components by derivatives of $\psi$ results in complex expressions 
that we shall not write out in full. We simply note, as an example, that the 
first lefthand term in the $r$-component of Euler's equation is
\begin{equation}
\rho^3\,u_r\,{{\partial u_r}\over{\partial r}} \,=\, 
-{{2\,\rho}\over{r^5\,{\rm sin}^2\,\theta}}
\left({{\partial\psi}\over{\partial\theta}}\right)^2 \,-\,
{1\over{r^4\,{\rm sin}^2\,\theta}}\,
{{\partial\rho}\over{\partial r}}
\left({{\partial\psi}\over{\partial\theta}}\right)^2 \,+\,
{1\over{r^4\,{\rm sin}^2\,\theta}}\,
{{\partial\psi}\over{\partial\theta}}\,
{{\partial^2 \psi}\over{\partial r \partial\theta}} \,\,.
\end{equation}
After substituting the series expansions for $\psi$ and $\rho$, we find that 
the highest power of $r$ is $r^{-1}$. In this case, all the coefficients on
both sides of Euler's equations vanish identically.

Matching the coefficients of $r^{-2}$, we obtain the {\it first-order}
equations. From the $r$-component of Euler's equation, we find
\begin{equation}\label{eqn:firstr}
-\beta\,f_1^{\prime\prime} \,-\, \beta\,f_1 \,+\,
\beta^2\,{\rm sin}\,\theta\,{\rm cos}\,\theta\,\,g_{-1}^\prime \,+\,
\left(\beta^2\,{\rm cos}^2\,\theta\,\,-\,1\right)\,g_{-1} \,+\,1 \,=\,0 \,\,,
\end{equation}
while the $\theta$-component yields
\begin{equation}\label{eqn:firstt}
\left(1\,-\,\beta^2\,{\rm sin}^2\,\theta\right)\,g_{-1}^\prime \,-\,
\beta^2\,{\rm sin}\,\theta\,{\rm cos}\,\theta\,\,g_{-1}\,=\,0 \,\,.
\end{equation}
In both of these equations and those that follow, a prime denotes a $\theta$-derivative.

These equations govern the first non-trivial terms in the expansions for $\psi$
and $\rho$. Their solution, therefore, must be equivalent to that obtained
through the more traditional, linear analysis. Integrating equation~(\ref{eqn:firstt}), we
find
\begin{equation}
g_{-1} \,=\, {C\over{\left(1\,-\,\beta^2\,{\rm sin}^2\,\theta\right)^{1/2}}}
\,\,,
\end{equation}
where $C$ is a constant, as yet undetermined. In the subsonic regime, the
denominator on the righthand side does not vanish for any $\theta$, and
$g_{-1}$ remains finite.

Substituting this expression for $g_{-1}$ into (\ref{eqn:firstr}) gives the equation obeyed
by $f_1$:
\begin{equation}
f_1^{\prime\prime} \,+\, f_1 \,=\, {1\over\beta} \,-\,
{{C\,\left(1\,-\,\beta^2\right)}\over
{\beta \left(1\,-\,\beta^2\,{\rm sin}^2\,\theta\right)^{3/2}}} \,\,.
\end{equation}
A particular solution of this equation may be found through the method of
variation of parameters. Adding the two homogeneous solutions yields
\begin{equation}
f_1 \,=\,{1\over\beta} \,-\,
{{C\,\left(1\,-\,\beta^2\,{\rm sin}^2\,\theta\right)^{1/2}}\over\beta} \,+\,
D\,{\rm cos}\,\theta \,+\, E\,{\rm sin}\,\theta \,\,,
\end{equation}
where $D$ and $E$ are also constants.

We proceed to evaluate the constants through application of the boundary 
conditions. The requirement that \hbox{$f_1 (\pi)\,=\,0$} gives
\begin{equation}
C\,=\,1\,-\,\beta\,D \,\,.
\end{equation}
Similarly, we have \hbox{$f_1 (0)\,=\,0$}, yielding
\begin{equation}
C\,=\,1\,+\,\beta\,D \,\,.
\end{equation}
It follows, from these last two relations, that \hbox{$C\,=\,1$} and
\hbox{$D\,=\,0$}. Finally, we have \hbox{$f_1^\prime (\pi)\,=\,0$}, from which
we infer that \hbox{$E\,=\,0$}. It may be verified that the boundary 
condition \hbox{$f_1^\prime (0) \,=\, 0$} is then also satisfied. 

In summary, the first-order density and stream function perturbations are
\begin{eqnarray}\label{eqn:gm1}
g_{-1} \,&=&\, 
{1\over{\left(1\,-\,\beta^2\,{\rm sin}^2\,\theta\right)^{1/2}}}\,\,, \\
\label{eqn:f1}
f_1 \,&=&\, {{1\,-\,\left(1\,-\,\beta^2\,{\rm sin}^2\,\theta\right)^{1/2}}\over
\beta} \,\,.
\end{eqnarray}
Notice that \hbox{$g_{-1}^\prime\,=\,0$} at both \hbox{$\theta\,=\,0$} and
$\pi$, implying that the density profile is flat (i.e., does not have a cusp)
on either the upstream or downstream axis. Our expression for $g_{-1}$ is
consistent with the linear density perturbation obtained by \citet{d64},
\citet{rs71}, and \citet{o99}. Notice that $g_{-1}(\pi/2)$ diverges as $\b$ approaches unity, signifying the birth of the Mach cone. Our $f_1$, in combination with $g_{-1}$, yields the linear velocity components given in equations~(25) and (26) of \citet{d64}. 

\begin{figure}
\plotone{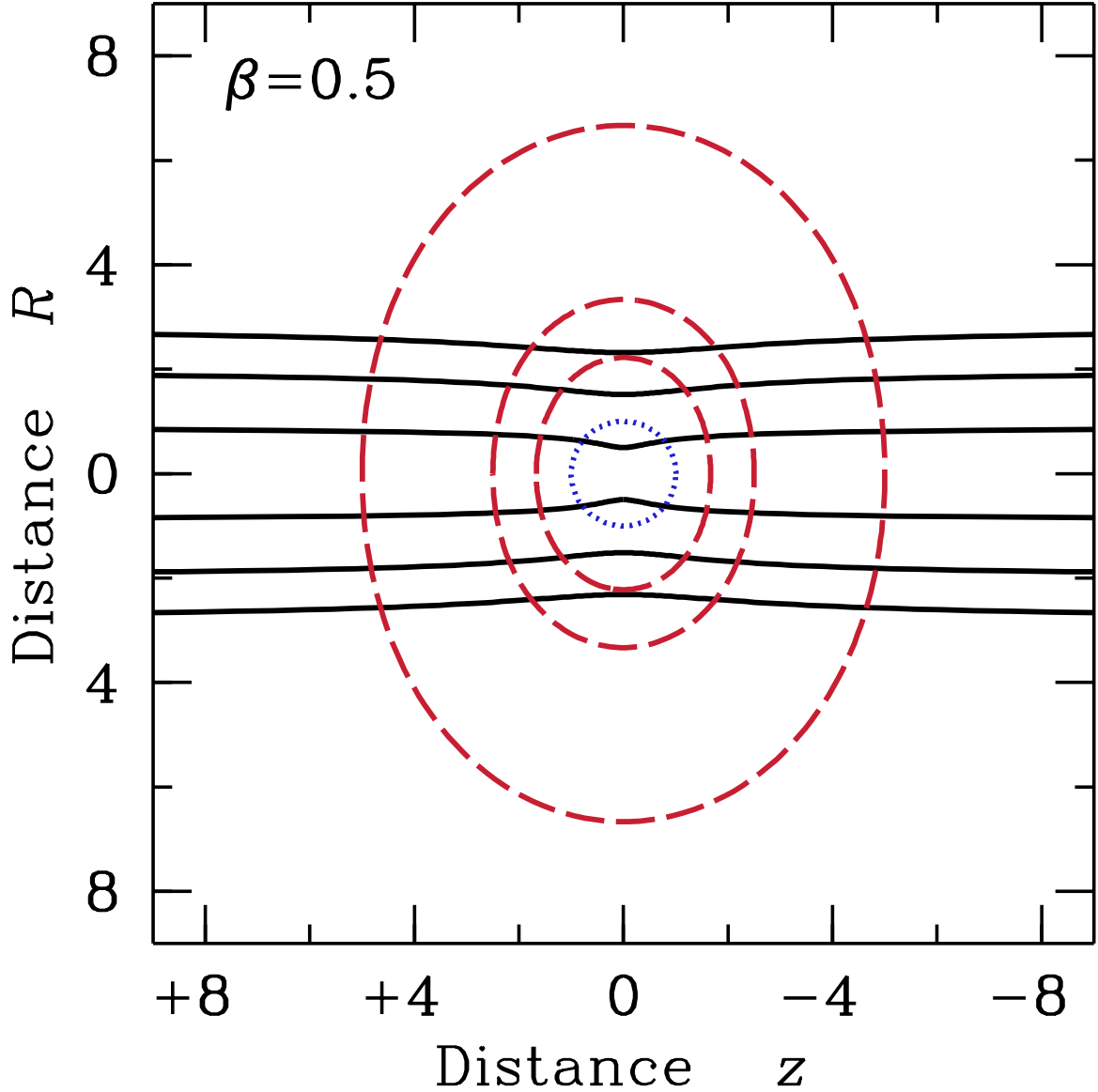}
\caption{Streamlines ({\it solid}) and density contours ({\it dashed}) for the $\b=0.5$ flow, including only first-order perturbations. All quantities shown are nondimensional. The density contours correspond to $\rho = 1.2, 1.4,$ and 1.6, while adjacent streamlines enclose equal mass fluxes. The inner dotted circle has the sonic radius. Since the streamlines and density contours are symmetric upstream and downstream, we cannot determine the true accretion rates of mass and momentum without higher-order perturbations. }
\label{fig:1storder}
\end{figure}

Figure~\ref{fig:1storder} displays the streamlines ({\it solid curves}) and isodensity contours
({\it dashed curves}) for the outer flow, including only the first-order
perturbations. The light, dotted circle marks the sonic radius, 
\hbox{$r\,=\,1$}; the solution is only accurate well outside this sphere.
Notice how all curves and contours are symmetric about the
\hbox{$\theta\,=\,\pi/2$} plane. The streamlines, in particular, show the 
fluid veering toward the mass, but then turning away again. We cannot detect 
true accretion of mass or linear momentum until we include the next higher-order perturbations. 

\subsection{Second-Order Equations}
We next equate coefficients of $r^{-3}$ in both components of Euler's equation.
From the $r$-component, equation~(\ref{eqn:eulerr}), we obtain one relation between $f_0$ and
$g_{-2}$:
\begin{equation}\label{eqn:secondr}
-\beta\,f_0^{\prime\prime} \,-\, \beta\,{\rm cot}\,\theta\,f_0^\prime \,+\,\
\beta^2\,{\rm sin}\,\theta\,{\rm cos}\,\theta\,\,g_{-2}^\prime \,+\,
\left(2\,\beta^2\,{\rm cos}^2\,\theta\,-\,2\right)\,g_{-2}\,=\,
{\cal A}_1 \,+\, {\cal A}_2 \,+\, {\cal A}_3 \,\,.
\end{equation}
Here, ${\cal A}_1$, ${\cal A}_2$, and ${\cal A}_3$ are expressions involving
$f_1$ and $g_{-1}$:
\begin{eqnarray}
{\cal A}_1 \,&\equiv&\, {f_1^2\over{{\rm sin}^2\,\theta}} \,-\,
{{f_1\,f_1^\prime\,{\rm cos}\,\theta}\over{{\rm sin}^3\,\theta}} \,+\,
{{\left(f_1^\prime\right)^2}\over{{\rm sin}^2\,\theta}} \,+\,
{{f_1\,f_1^{\prime\prime}}\over{{\rm sin}^2\,\theta}}\,\,,
\\ 
{\cal A}_2 \,&\equiv&\, \beta\,f_1\,\,g_{-1} \,-\,
2\,\beta\,f_1^\prime\,\,g_{-1}\,{\rm cot}\,\theta \,-\,
\beta\,f_1\,\,g_{-1}^\prime\,{\rm cot}\,\theta \,-\,
\beta\,f_1^\prime\,\,g_{-1}^\prime \,+\,
\beta\,f_1^{\prime\prime}\,\,g_{-1} \,\,,
\\
{\cal A}_3 \,&\equiv&\, 2\,g_{-1}^2  \,-\, 3\,g_{-1}
\,\,.
\end{eqnarray}
From the $\theta$-component, equation~(\ref{eqn:eulert}), we obtain a second relation between
$f_0$ and $g_{-2}$:
\begin{equation}\label{eqn:secondt}
-\beta\,f_0^\prime \,+\, {\cal D}\,g_{-2}^\prime \,-\,
2\,\beta^2\,{\rm sin}\,\theta\,\,{\rm cos}\,\theta\,\,g_{-2} \,=\,
{\cal B}_1 \,+\, {\cal B}_2 \,+\, {\cal B}_3
\,\,,
\end{equation}
where we have defined 
\hbox{${\cal D}\,\equiv\,1\,-\,\beta^2\,{\rm sin}^2\,\theta$}, and where the
three terms on the righthand side are again combinations of $f_1$ and $g_{-1}$:
\begin{eqnarray}
{\cal B}_1 \,&\equiv&\, f_1^2\,\frac{{\rm cot}\,\theta}{\sin^2\t} \,-\,
{{f_1\,f_1^\prime}\over{{\rm sin}^2\,\theta}} \,\,,
\\
{\cal B}_2 \,&\equiv&\, \beta\,f_1\,\,g_{-1}\,{\rm cot}\,\theta \,\,+\,\,
\beta\,f_1^\prime\,\,g_{-1} \,\,+\,\,2\,\beta\,f_1\,g_{-1}^\prime \,\,,
\\
{\cal B}_3 \,&\equiv&\, -2\,g_{-1}\,\,g_{-1}^\prime
\,\,.
\end{eqnarray}

We have already found $f_1$ and $g_{-1}$ in the subsonic case of interest. 
After substituting these expressions, equations~(\ref{eqn:gm1}) and (\ref{eqn:f1}), into the 
righthand sides of equations~(\ref{eqn:secondr}) and (\ref{eqn:secondt}), the coupled equations for $f_0$
and $g_{-2}$ become:
\begin{equation} \label{eqn:secondr2}
-\beta\,f_0^{\prime\prime} \,-\, \beta\,{\rm cot}\,\theta\,f_0^\prime \,+\,\
\beta^2\,{\rm sin}\,\theta\,{\rm cos}\,\theta\,\,g_{-2}^\prime \,+\,
\left(2\,\beta^2\,{\rm cos}^2\,\theta\,-\,2\right)\,g_{-2}\,=\,
{1\over{\cal D}}\,-\,{3\over\sqrt{\cal D}} \,+\,
{2\over{1+\sqrt{\cal D}}}\,\,,
\end{equation} 
and
\begin{equation}\label{eqn:secondt2}
-\beta\,f_0^\prime \,+\, {\cal D}\,g_{-2}^\prime \,-\,
2\,\beta^2\,{\rm sin}\,\theta\,\,{\rm cos}\,\theta\,\,g_{-2} \,=\,
\beta^2\,{\rm sin}\,\theta\,\,{\rm cos}\,\theta
\left[-{2\over{{\cal D}^2}} \,+\, {2\over{{\cal D}^{3/2}}}\,-\,
{1\over{\cal D}} \,+\, {1\over{\left(1+\sqrt{\cal D}\right)^2}}
\right]\,\,.
\end{equation}

These last two relations constitute our {\it second-order} equations. For any
value of $\beta$, we may integrate them numerically from the upstream axis, 
\hbox{$\theta\,=\,\pi$}, to the downstream axis at \hbox{$\theta\,=\,0$}. Three
initial conditions are required, of which we have already identified two:
\hbox{$f_0 (\pi) \,=\, f_0^\prime (\pi) \,=\, 0$}. As a third initial 
condition, we use $g_{-2} (\pi)$, whose value at this point is arbitrary. For
each chosen value of $g_{-2} (\pi)$, we may find $f_0 (\theta)$ and
$g_{-2} (\theta)$. We thus have a one-parameter family of outer flow solutions.

\begin{figure}
\plotone{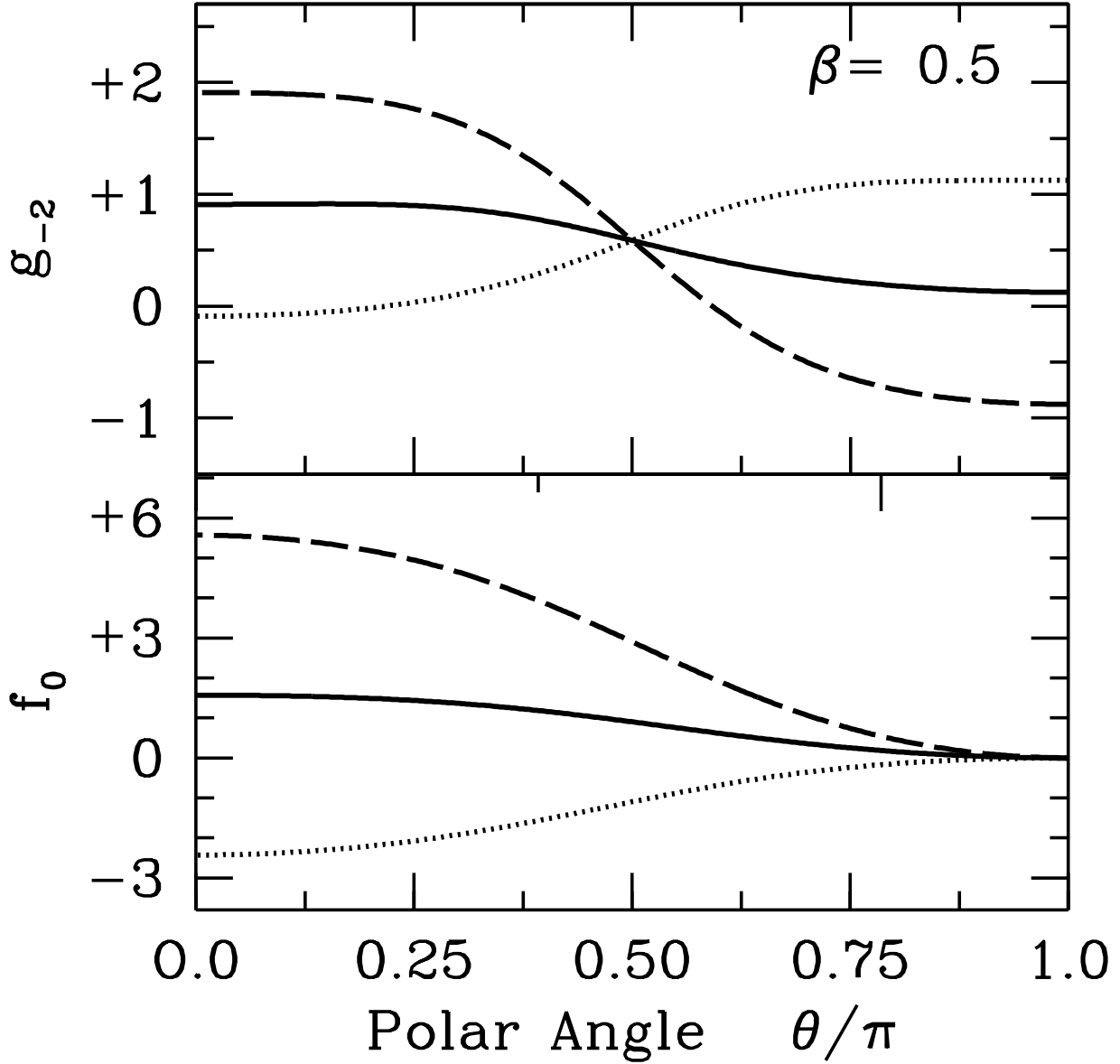}
\caption{Sample solutions to the second-order perturbation equations (\ref{eqn:secondr2}) and (\ref{eqn:secondt2}), for $g_{-2}(\theta)$ ({\it upper panel}) and $f_0(\theta)$ ({\it lower panel}). The initial values of $g_{-2}(\pi)$ are +1.12 ({\it dotted}), +0.12 ({\it solid}), and -0.88 ({\it dashed}). }
\label{fig:2ndfandg}
\end{figure}

In the upper panel of Figure~\ref{fig:2ndfandg} we display, for the representative value 
\hbox{$\beta\,=\,0.5$}, three solutions of $g_{-2} (\theta)$. We obtained each
solution by assuming different values of $g_{-2} (\pi)$. Notice that $g_{-2}^\prime$ vanishes on the upstream and downstream axes, implying again that the density profile is flat in both regions. Notice also that all curves attain the same value at \hbox{$\theta\,=\,\pi/2$}. That is, $g_{-2} (\pi/2)$ depends only on $\beta$, 
and not on the prescribed initial condition $g_{-2} (\pi)$.

The lower panel of Figure~\ref{fig:2ndfandg} shows the corresponding plots of $f_0 (\t)$. We see that 
\hbox{$f_0^\prime (0) \,=\,0$} in every case, ensuring regularity of $u_r$ on 
the downstream axis. This condition was not imposed a priori, but resulted 
automatically from integration of the governing equations. Specifically, the coefficient of $f_0^\prime$ in equation~(\ref{eqn:secondr2}) includes 
${\rm cot}\,\theta$, which diverges at \hbox{$\theta\,=\,0$}. Since all the 
other terms in this equation remain finite on the axis, $f_0^\prime (0)$ is 
forced to zero.

Which of these solutions is the true outer flow for gas that is accreting
steadily onto the gravitating mass? In principle, one could answer this 
question by continuing each solution inward, to see if the flow smoothly
crosses the sonic surface, where \hbox{$u\,=\,1$}. We shall not attempt such
a calculation here. Instead, we will proceed by determining generically
the mass accretion rate that is associated with each outer solution. Then,
given the Bondi prescription for this rate, we will indeed be able to select
the physical solution for each $\beta$. 

\section{Mass Accretion Rate}
\label{sec:mass}
\subsection{Relation to Stream Function}
One could, in principle, equate coefficients of $r^{-4}$, $r^{-5}$, etc., and
thereby obtain the coupled equations linking higher-order $f$- and 
$g$-variables. We will now demonstrate, however, that the first- and 
second-order equations just presented are sufficient to establish the total 
accretion rate onto the mass. We will then relate, in Section~\ref{sec:friction} below, this 
infall rate to the desired friction force.  

Refer again to Figure~\ref{fig:coord} and imagine a sphere of radius $r$ surrounding the mass. Reverting temporarily to 
dimensional variables, the mass accretion rate is
\begin{eqnarray}
{\dot M} \,&=&\, -2\,\pi\int_0^\pi\!\rho\,u_r\,r^2\,{\rm sin}\,
           \theta\,d\theta \\
\,&=&\, -2\,\pi\int_0^\pi\!
{{\partial\psi}\over{\partial\theta}}\,d\theta 
\\
\,&=&\, 2\,\pi\,\left[\psi (r,0)\,-\,\psi (r,\pi)\right]
\,\,,
\end{eqnarray}
where we have utilized equation~(\ref{eqn:ur}) connecting $u_r$ and $\psi$. Recall that
$\psi (r,\pi)$ is actually a constant, independent of $r$, and that we have
set that constant to zero. We thus have
\begin{equation}
{\dot M} \,=\, 2\,\pi\,\psi(r,0) \,\,.
\end{equation}

To nondimensionalize this result, we first set the fiducial mass accretion rate
to \hbox{$2\,\pi\,\rho_0\,c_s\,r_s^2$}. After using the expansion of $\psi$ 
from equation~(\ref{eqn:psind}), we obtain the nondimensional equation
\begin{equation}
{\dot M} \,=\, f_2 (0)\,r^2 \,+\, f_1 (0)\,r \,+\, f_0 (0) \,+\,
f_{-1}(0)\,r^{-1} \,+\, f_{-2} (0)\,r^{-2} \,\,...  \,\,.
\end{equation}
One of our boundary conditions, ensuring regularity of $u_\theta$ on the
downstream axis, is that \hbox{$f_i (0)\,=\,0$} for 
\hbox{$i\,=\,1,\,-1,\,-2,\,$} etc. Since \hbox{$f_2 (0)\,=\,0$}, we find
the simple relation
\begin{equation}\label{eqn:mdotf0}
{\dot M} \,=\, f_0 (0) \,\,.
\end{equation}

Both sides in this equation are functions of $\beta$, although we have not
indicated the dependence explicitly. In any case, the relation confirms our
expectation that the mass accretion rate is independent of the sphere's radius
$r$ in steady-state motion.\footnote{Note, however, that the original series
expansion for $\psi$ becomes inaccurate when $r$ is not much greater than 
unity.} We also now see that the higher-order variables $f_{-1}$,
$f_{-2}$, etc. play no part in determining this rate.

\subsection{Relation to Density Perturbation}
Now that we have tied the mass accretion rate to $f_0 (0)$, we can immediately
rule out a subset of outer flow solutions as being unphysical. Figure~\ref{fig:2ndfandg} shows
that, for \hbox{$g_{-2} (\pi)\,=\,1.12$}, $f_0 (0)$ is negative, 
corresponding to a net mass efflux. That such a situation is even possible 
emphasizes once more the need to extend the flow solution inward across the 
sonic surface.

For this same choice of $g_{-2} (\pi)$, the dotted curve in the lower panel of
Figure~\ref{fig:2ndfandg} shows that
\hbox{$g_{-2}(0)\,<\,g_{-2}(\pi)$}. Indeed, we have just found one example 
of a general result: the difference 
\hbox{$g_{-2} (0) \,-\, g_{-2} (\pi)$} agrees in sign with $f_0 (0)$. We now 
show that the two quantities are in fact equal, apart from a multiplicative 
factor.

Our proof starts with the fact that the lefthand side of the second-order
equation~(\ref{eqn:secondt2}) is a perfect derivative. Specifically,
\begin{equation}
-\beta\,f_0^\prime \,+\, {\cal D}\,g_{-2}^\prime \,-\,
2\,\beta^2\,{\rm sin}\,\theta\,\,{\rm cos}\,\theta\,\,g_{-2} \,\,=\,\,
{{d{\phantom\theta}}\over{d\,\theta}}
\left(-\beta\,f_0 \,+\,{\cal D}\,g_{-2}\right) \,\,.
\end{equation}
Turning to the righthand side of the same equation, we note first that
${\rm sin}\,\theta$ is an even function of $\theta-\pi/2$, while
${\rm cos}\,\theta$ is an odd function. Since $\cal D$ depends only on
${\rm sin}\,\theta$, it has even symmetry. Inspection shows that the
righthand side of equation~(\ref{eqn:secondt2}) has odd symmetry.

If we now integrate equation~(\ref{eqn:secondt2}) from \hbox{$\theta\,=\,\pi$} to 0, the
righthand side vanishes because of the odd symmetry of the integrand. We find
that
\begin{equation}
\left(-\beta\,f_0 \,+\,{\cal D}\,g_{-2}\right)_{\theta\,=\,\pi} \,=\,
\left(-\beta\,f_0 \,+\,{\cal D}\,g_{-2}\right)_{\theta\,=\,0} \,\,.
\end{equation} 
Since \hbox{$f_0 (\pi) \,=\, 0$} and 
\hbox{${\cal D} (\pi) \,=\, {\cal D} (0) \,=\, 1$}, we have
\begin{equation}
g_{-2} (\pi) \,=\, -\beta\,f_0 (0) \,+\, g_{-2} (0) \,\,,
\end{equation} 
which we recast as
\begin{equation}
f_0 (0) \,=\,
{{g_{-2} (0) \,-\, g_{-2} (\pi) }\over\beta} \,\,.
\end{equation} 
Recalling equation~(\ref{eqn:mdotf0}) that identifies $f_0 (0)$ as the mass accretion rate, 
we now see that this rate is proportional to the difference, upstream and 
downstream, of the second-order density perturbation. As $\beta$ approaches 
zero, these two perturbations become equal. Indeed, the function 
$g_{-2} (\theta)$ is a constant (equal to 1/2) in the limit, consistent
with a spherically symmetric flow. 

\subsection{Modified Bondi Prescription}
To establish the physically relevant flow solutions, we need to specify the
accretion rate as a function of velocity. \citet{b52} fully solved the 
\hbox{$\beta \,=\, 0$} problem. That is, he determined the complete 
distribution of density and velocity surrounding a mass at rest within a 
background gas. Dimensionally, he found for the mass accretion rate
\begin{equation}
{\dot M} \,=\, {{4\,\pi\,\lambda\,\rho_0\,G^2\,M^2}\over {c_s^3}} \,\,,
\end{equation}
where \hbox{$\lambda\,=\,{\rm e}^{3/2}/4 \,=\, 1.12$} for the isothermal case
of interest here. Recasting the rate into nondimensional form (recall Section 4.1), we have
\begin{eqnarray}
\lim_{\beta\,\rightarrow\,0} \,\,
{\dot M}  \,&=&\, 2\,\lambda  \nonumber  \\
\label{eqn:bondimdot}
\,&=&\, {{{\rm e}^{3/2}}\over 2} \,\,,
\end{eqnarray}
as one constraint on the general form of $\dot M (\beta)$. 

Prior to Bondi's work, \citet{hl39} studied accretion onto a mass traveling
through a zero-temperature gas. Their dimensional result was 
\begin{equation}
{\dot M} \,=\, {{4\,\pi\,\rho_0\,G^2\,M^2}\over{V^3}} \,\,.
\end{equation}
Noting that the Mach number $\beta$ is effectively infinite in this case, the
equivalent, nondimensional finding is 
\begin{equation}
\lim_{\beta\,\rightarrow\,\infty}  \,\,
{\dot M} \,=\,{2\over{\beta^3}} \,\,.
\end{equation}
\citet{bh44} later showed that this relation provides an upper bound to the
accretion rate in the zero-temperature case. Through more careful analysis of
the wake, which here degenerates into an infinite-density spindle, they set
the lower limit a factor of two smaller. 

The widely used interpolation formula of \citet{b52} connects these limits, at 
least approximately. Nondimensionally, the Bondi prescription is  
\begin{equation}
{\dot M} (\beta) \,=\,{1\over{\left(1\,+\,\beta^2\right)^{3/2}}} \,\,.
\end{equation}
In the low-$\beta$ limit, $\dot M$ falls short of the isothermal result, but
matches that for a \hbox{$\gamma\,=\,3/2$} polytrope. The high-$\beta$ limit
reproduces the lower bound established by \citet{bh44}.

Since we are focusing on the subsonic regime within an isothermal gas, we want
our low-$\beta$ limit to agree with the exact result. Following \citet{mt09},
we adopt a modified form of the classic interpolation formula:
\begin{equation}\label{eqn:intermdot}
{\dot M} (\beta) \,=\, {{2\,\left(\lambda^2\,+\,\beta^2\right)^{1/2}}\over 
{\left(1\,+\,\beta^2\right)^2}} \,\,,
\end{equation}
where we use the isothermal value of $\lambda$ previously given. For
\hbox{$\beta\,\ll\,1$}, $\dot M$ approaches the result of \citet{b52} given in
equation~(\ref{eqn:bondimdot}). For \hbox{$\beta\,\gg\,1$}, we recover the upper limit of 
\citet{bh44}. In the simulation of \citet{mt09} for an isothermal gas with $\beta=10$, this 
modified interpolation formula matches the calculated accretion rate to within 
20~percent. Judging from their own polytropic simulations, both \citet{h71} and 
\citet{s85} had earlier suggested that the original Bondi $\dot M (\beta)$ be 
augmented by about a factor of two. 
In summary, equation~(\ref{eqn:intermdot}) should be 
sufficiently accurate for our purposes. 

\begin{figure}
\plotone{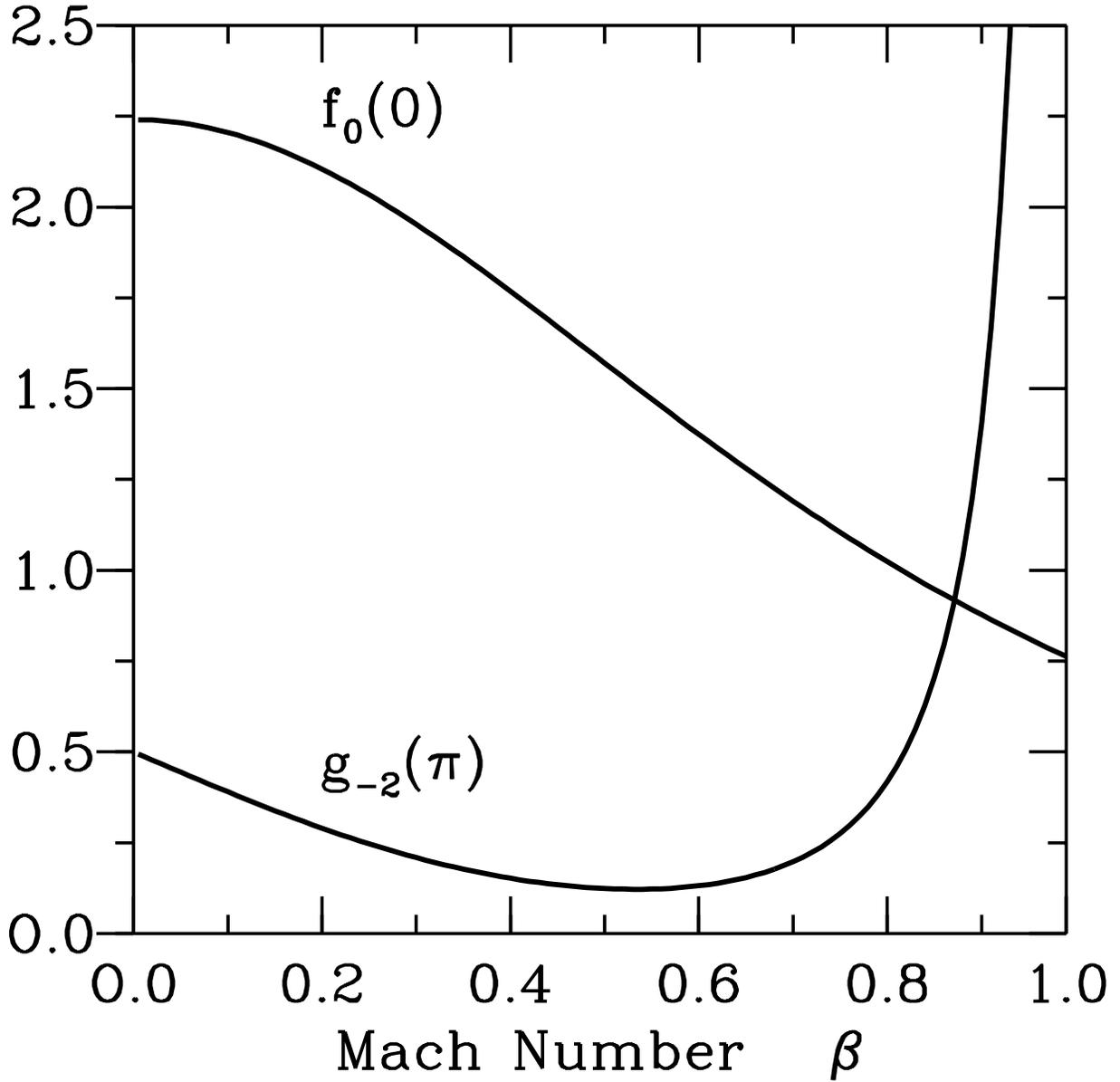}
\caption{The upstream density perturbation $g_{-2}(\pi)$ for the physical accretion flow, shown as a function of Mach number $\b$. This initial condition gives the correct $\dot{M} = f_0(0)$, also shown in the figure.}
\label{fig:2ndg}
\end{figure}

The combination of equations~(\ref{eqn:mdotf0}) and (\ref{eqn:intermdot}) gives us the proper value of
$f_0 (0)$ at each $\beta$, and thus also establishes the physically relevant
outer flow solutions. Figure~\ref{fig:2ndg} shows the physical $f_0 (0)$ and $g_{-2}(\pi)$ as
functions of $\beta$. Note that the latter diverges as $\b$ approaches unity. Thus, our perturbation series fails to describe the flow along the upstream axis in this limit. As we will show in the next section, however, the dynamical friction force remains finite for all $\b$. 

\begin{figure}
\vspace{-0.45in}

\begin{center} 
	\includegraphics[scale=0.4]{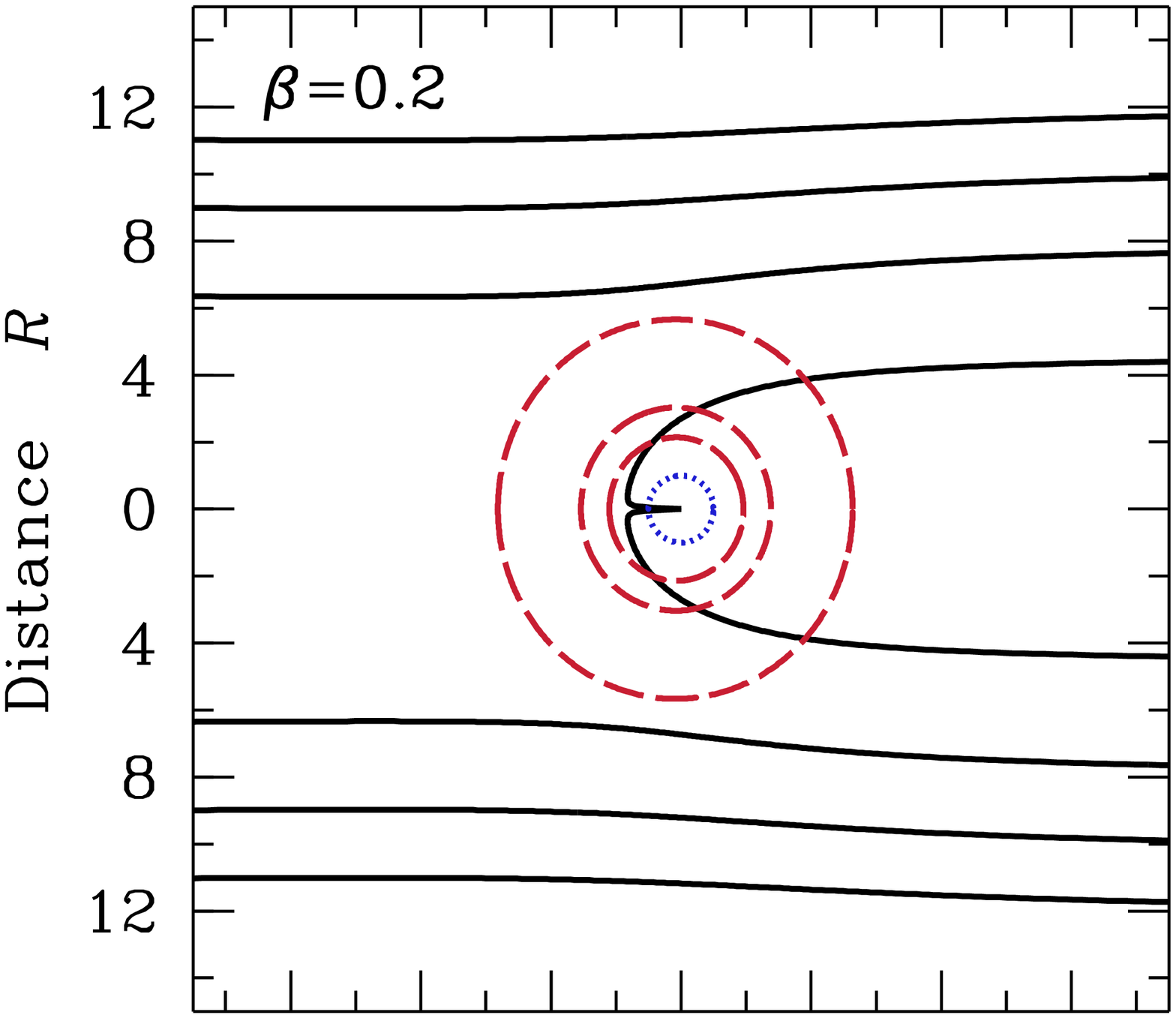}\vspace{-0.63in}
	
	\includegraphics[scale=0.4]{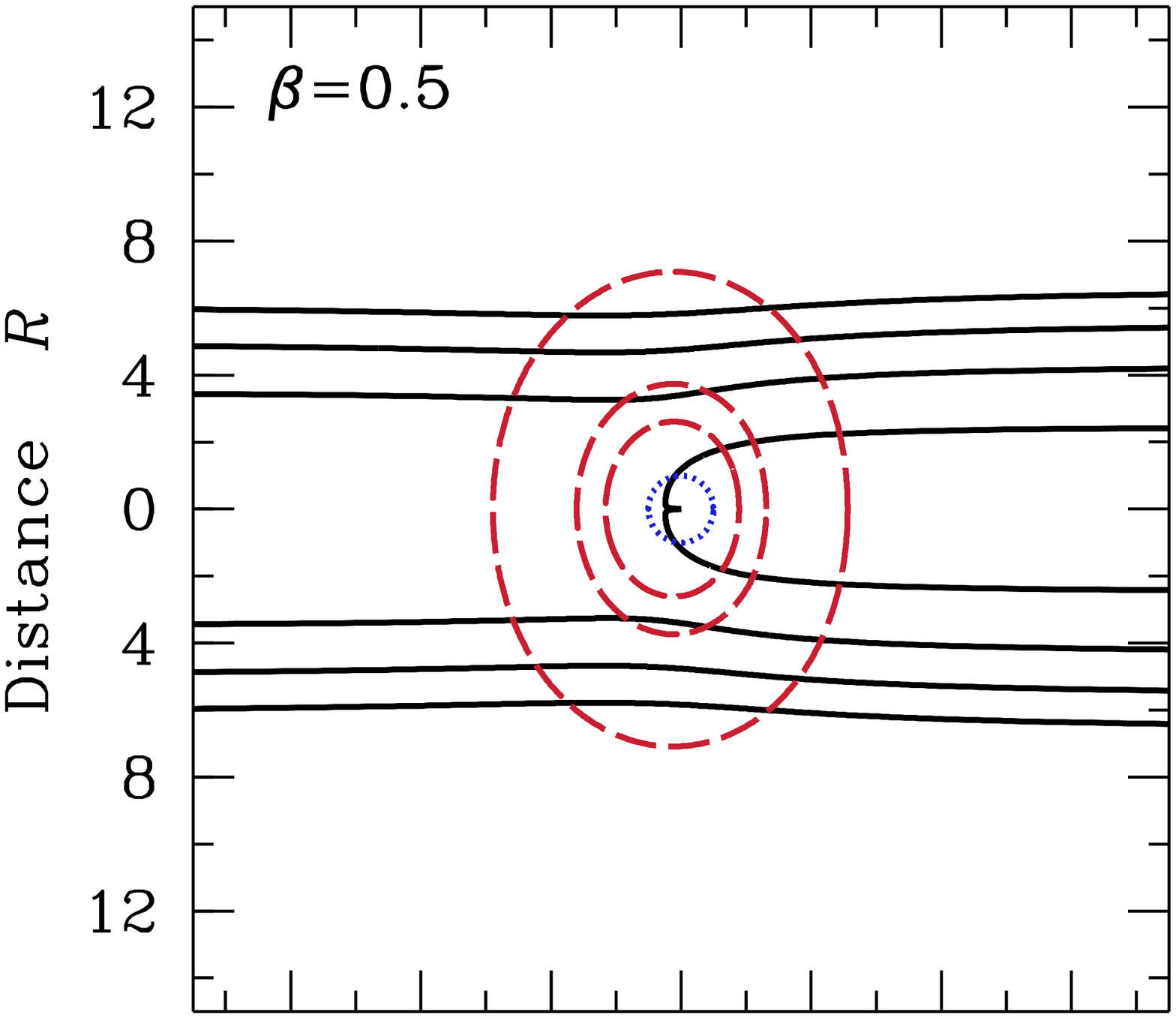}\vspace{-0.63in}
	
	\includegraphics[scale=0.4]{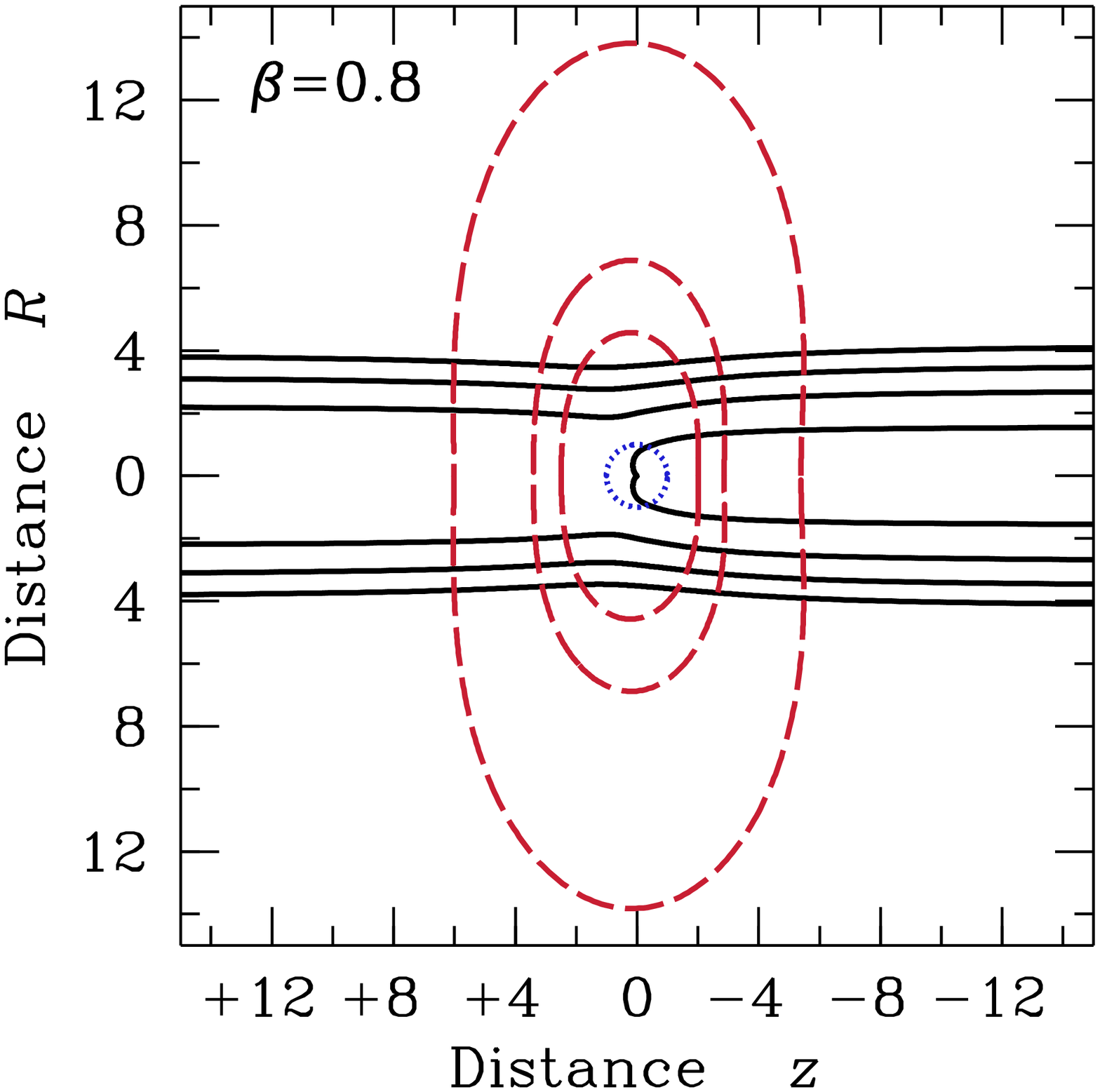} 
\end{center}
\caption{Streamlines ({\it solid}) and density contours ({\it dashed}) for three different Mach numbers $\beta$. The density contours correspond to $\rho=1.2,1.4,$ and 1.6. The innermost streamlines enclose the full mass accretion rate $\dot{M}$. Successive streamlines enclose 3, 5, and 7 times this rate. As in Figure~\ref{fig:1storder}, the inner circle represents the sonic surface. }
\label{fig:2ndstream}
\end{figure}

The three panels of Figure~\ref{fig:2ndstream} display streamlines and 
isodensity contours for the indicated $\beta$-values. These curves were 
constructed from equations~(\ref{eqn:psind}) and (\ref{eqn:rhond}) for $\psi$ and $\rho$, respectively, 
using the three known terms in each series. The circle in each panel represents
the sonic surface. As always, our results are only accurate well beyond this 
radius.

\section{Friction Force}
\label{sec:friction}
\subsection{Integral Expression}
The dynamical friction force $F$ is the total rate at which $z$-momentum is
transferred from the background gas to the gravitating mass. Within our
steady-state flow, the total momentum transfer rate into a surface surrounding
the mass is independent of the size and shape of that surface, provided it 
lies outside the wake, where the physical interaction between the projectile 
and gas occurs. The net momentum flow calculated through such a surface 
integration all goes into the gravitating mass, causing its deceleration. 

Imagine the gravitating mass to be surrounded by a large sphere of 
radius $r$. In part, the $z$-momentum transfer arises from the advection of 
this quantity in the flowing gas across the spherical surface. Since the inward
flux of $z$-momentum is $-\rho\,u_r\,u_z$ the {\it kinetic} portion of $F$ is, 
dimensionally,
\begin{equation}
F_{\rm kin} \,=\, -2\,\pi \,\int_0^\pi\!
              \rho\,u_r\,u_z\,r^2\,{\rm sin}\,\theta\,d\theta \,\,.
\end{equation} 
Another contribution to $F$ is from the thermal pressure of the surrounding
gas. This {\it static} portion of the force is
\begin{equation}
F_{\rm static} \,=\, -2\,\pi\,\int_0^\pi\!
                 \rho\,c_s^2\,r^2\,{\rm cos}\,\theta\,{\rm sin}\,\theta\,
                 d\theta \,\,.
\end{equation}
Adding these two pieces, we have, after nondimensionalization, 
\begin{equation}
F \,=\, -\int_0^\pi\!\rho\,u_r\,u_z\,r^2\,{\rm sin}\,\theta\,d\theta \,-\,
        \int_0^\pi\!\rho\,r^2\,{\rm cos}\,\theta\,{\rm sin}\,\theta\,
        d\theta \,\,,
\end{equation} 
where we have set the unit of force equal to 
\hbox{$2\,\pi\,\rho_0\,c_s^2\,r_s^2$}.  

The integrand within the first, righthand term must be recast in terms of the
stream function:
\begin{equation}
\rho\,u_r\,u_z\,r^2\,{\rm sin}\,\theta \,=\,
{{{\rm cot}\,\theta}\over{\rho\,r^2}} 
\left({{\partial\psi}\over{\partial\theta}}\right)^2 \,+\,
{1\over{\rho\,r}} 
{{\partial\psi}\over{\partial\theta}}
{{\partial\psi}\over{\partial r}} \,\,.
\end{equation} 
We may now evaluate $F$ using the series expansions for $\psi$ and $\rho$. The
full expression is a series of terms proportional to $r^2$, $r^1$, $r^0$, etc.

All terms in $F$ containing positive powers of $r$ vanish upon integration. 
Those proportional to $r^1$ involve $f_1$ and $g_{-1}$, both of 
which are known explicitly. Terms associated with negative powers of $r$
contain $f$- and $g$-variables which we have not yet calculated
(e.g., $f_{-1}$, $g_{-3}$). However, as we consider ever larger radii $r$, 
where the series expansions themselves become increasingly accurate, these
terms also go to zero. Only those independent of $r$ survive.

After restricting ourselves to $r$-independent terms, we find 
\begin{equation}\label{eqn:friction}
F \,=\, -\int_0^\pi\! \left[
        \left(1-\beta^2\right)\,{\rm sin}\,\theta\,
        {\rm cos}\,\theta\,\,g_{-2}
        \,+\,\beta \left(1\,+\,{\rm cos}^2\,\theta\right)
        f_0^\prime
        \right]\! d\theta \,\,.
\end{equation}
Here, we have omitted a number of terms in the integrand containing $f_1$, 
$g_{-1}$, and their derivatives. All of these terms are antisymmetric
with respect to \hbox{$\theta\,-\pi/2$} (i.e., they are odd functions), and
therefore vanish upon integration.

\subsection{Relation to Mass Accretion Rate}

By dimensional considerations, the friction force should be 
\hbox{$F\,=\,C{\dot M}\,V$}, where $C$ is dimensionless. In the hypersonic limit, this multiplicative factor contains a Coulomb logarithm 
\citep[e.g.][]{rs71}. We now demonstrate the surprising fact that, in the 
subsonic case of interest here, the factor is exactly unity. In fully nondimensional
language, we shall prove that
\begin{equation}\label{eqn:mdotv}
F\,=\,{\dot M}\,\beta \,\,.
\end{equation}

We begin by splitting the integral on the righthand side of equation~(\ref{eqn:friction}) into
two parts:
\begin{mathletters}
\begin{eqnarray}
F\,&=&\, -\int_0^\pi\! \beta\,f_0^\prime\,d\theta \,-\,
       \int_0^\pi\! \left[
       \left(1-\beta^2\right)\,{\rm sin}\,\theta\,
       {\rm cos}\,\theta\,\,g_{-2}
       \,+\,\beta\,{\rm cos}^2\,\theta\,
       f_0^\prime
       \right]\! d\theta \\
{\phantom F} \,&=&\, \beta\,f_0 (0) \,-\, {\cal I} \,\, \\
{\phantom F} \,&=&\, \beta\,\dot{M} \,-\, {\cal I} \,\,.
\end{eqnarray}
\end{mathletters}
In these equations, we have used the
fact that \hbox{$f_0 (\pi)\,=\,0$} and $f_0(0)=\dot{M}$ (eq. \ref{eqn:mdotf0}). We have further defined
\begin{equation}
{\cal I} \,\equiv\, 
       \int_0^\pi\! d\theta\,\left[
       \left(1-\beta^2\right)\,{\rm sin}\,\theta\,
       {\rm cos}\,\theta\,\,g_{-2}
       \,+\,\beta\,{\rm cos}^2\,\theta\,
       f_0^\prime
       \right] \,\,.
\end{equation}
We next show that $\cal I$ vanishes.

First recall that our flow is irrotational. Specifically, the $\phi$-component of the vorticity vanishes, so that
\begin{equation}
\frac{\partial\,u_r}{\partial\,\theta} \,-\,
\frac{\left(r\,u_\theta\right)}{\partial\,r} \,=\,0 \,\,.
\end{equation}
Expressing both velocity components in terms of the stream function through 
equations~(\ref{eqn:ur}) and (\ref{eqn:ut}), we have
\begin{equation}
\frac{\partial\rho}{\partial\theta}\,\frac{\partial\psi}{\partial\theta}\,+\,
\rho\,{\rm cot}\,\theta\,\frac{\partial\psi}{\partial\theta} \,-\,
\rho\,\frac{\partial^2\psi}{\partial \theta^2}\,+\,
r^2\,\frac{\partial\rho}{\partial r}\,\frac{\partial\psi}{\partial r}\,-\,
\rho\,r^2\,\frac{\partial^2\psi}{\partial r^2} \,=\, 0 \,\,.
\end{equation}

We substitute the series expansions for $\psi$ and $\rho$ into this last
equation and set the coefficients of all powers of $r$ to zero. Following this 
procedure for $r^2$ and $r^1$, and using the known expressions for $f_2$, 
$f_1$, and  $g_{-1}$, yields identities. However, setting the $r$-independent 
terms to zero leads to a nontrivial result: 
\begin{equation}
\beta\,f_0^{\prime\prime} \,-\,\beta\,{\rm cot}\theta\,f_0^\prime \,-\,
\beta^2\,{\rm sin}\,\theta\,{\rm cos}\,\theta\,g_{-2}^\prime\,+\,
2\,\beta^2\,{\rm sin}^2\,\theta\,g_{-2} \,=\,
\frac{1\,-\,\sqrt{\cal D}}{\cal D} \,\,.
\end{equation}

We add this last equation to the second-order equation~(\ref{eqn:secondr2}), obtaining
\begin{equation}
2\,\left(\beta^2\,-\,1\right) g_{-2} \,-\,   
2\,\beta\,{\rm cot}\,\theta\,f_0^\prime 
\,=\,\frac{2}{\cal D} \,-\,
\frac{4}{\sqrt{\cal D}} \,+\,
\frac{2}{1\,+\,\sqrt{\cal D}} \,\,.
\end{equation}
Multiplying through by \hbox{$-(1/2)\,{\rm sin}\,\theta\,{\rm cos}\,\theta$}
gives
\begin{equation}
\left(1\,-\,\beta^2\right) {\rm sin}\,\theta \,{\rm cos}\,\theta \,g_{-2} \,+\,
\beta\,{\rm cos}^2\,\theta\,f_0^\prime  \,=\,  
-{\rm sin}\,\theta\,{\rm cos}\,\theta
\left(\frac{1}{\cal D} \,-\,
\frac{2}{\sqrt{\cal D}} \,+\,
\frac{1}{1\,+\,\sqrt{\cal D}} \right)
\,\,.
\end{equation}
Integrating over $\theta$, we recognize the lefthand side of the resulting 
equation as $\cal I$. The righthand side vanishes, since the integrand is an 
odd function. We see therefore that equation (\ref{eqn:mdotv}) holds.  

\begin{figure}
\plotone{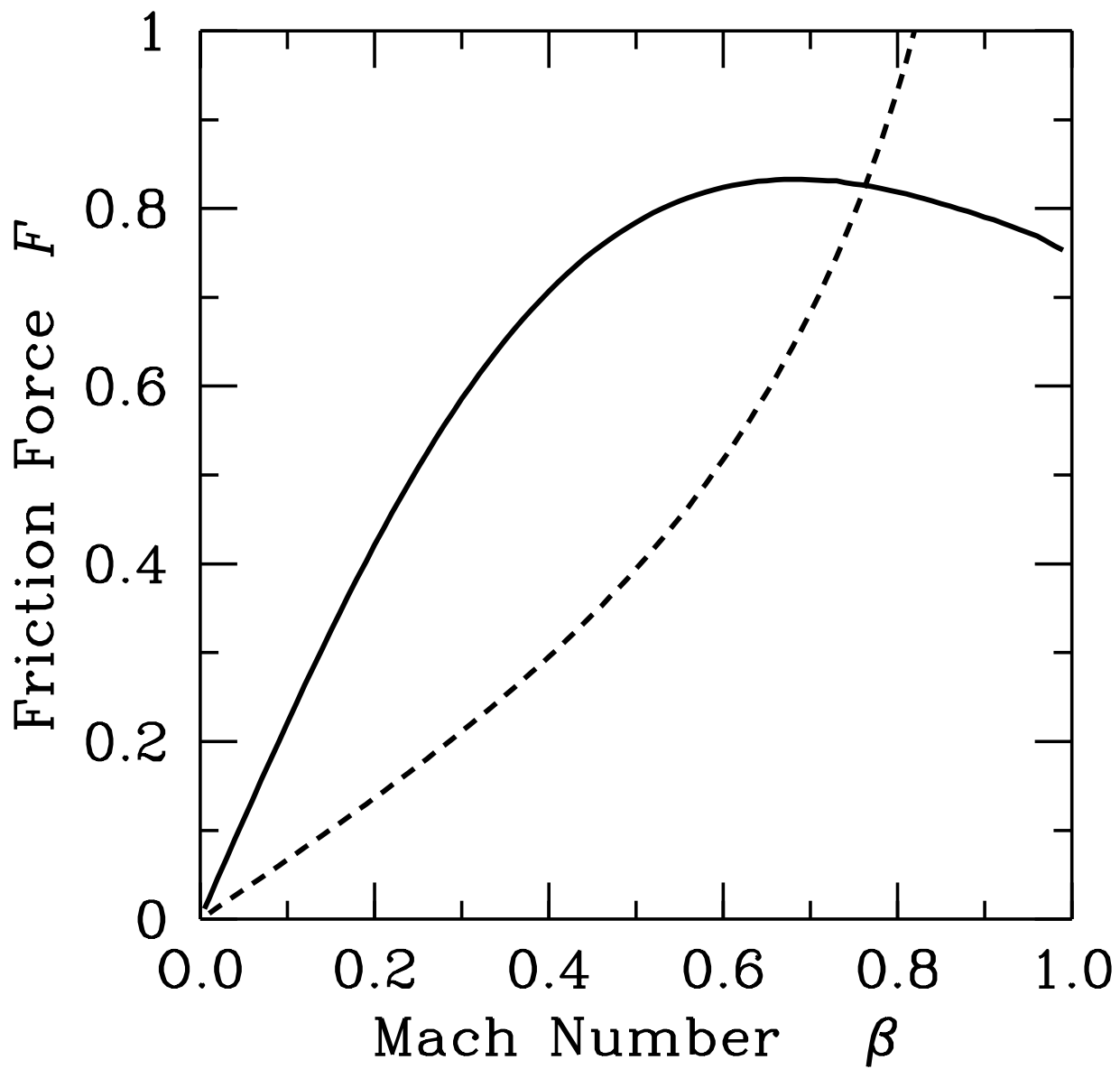}
\caption{The dimensionless friction force $F$ as a function of Mach number $\beta$. The  dashed curve shows the force derived by \citet{o99}, which diverges as $\b$ approaches unity.  
}
\label{fig:friction}
\end{figure}

If we now employ the modified Bondi prescription, equation~(\ref{eqn:intermdot}) for $\dot M$,
we have an explicit expression for the force:
\begin{equation}\label{eqn:bondifric}
F \,=\, {{2\,\beta\,\left(\lambda^2\,+\,\beta^2\right)^{1/2}}\over 
{\left(1\,+\,\beta^2\right)^2}} \,\,.
\end{equation}
Figure~\ref{fig:friction} displays the function $F (\beta)$. Also shown, as the dashed curve, is
the result from \citet{o99} in which the force diverges as $\b$ approaches unity. In the limit of low $\b$, both forces rise linearly from zero at $\b=0$, but our initial slope is larger by a factor of $3\lambda=3.36$. Indeed, over most $\b$-values, our force exceeds that derived by \citet{o99}, presumably because we have included both the gravitational tug from the wake and the direct accretion of momentum from the flow. Our force does not rise monotonically but instead peaks around $\b=0.68$ and then begins to decline; we expect this decline to continue into the supersonic regime. We should bear in mind that, while equation~(\ref{eqn:mdotv}) is exact, equation~(\ref{eqn:bondifric}) for $F$ is only as accurate as the underlying interpolation formula.

\section{Velocity and Mass Evolution}
\label{sec:velocity}
Our simple result for the dynamical friction force means that the 
deceleration of the gravitating mass is also simply described, as long
as there are no other forces at play. As we have stressed, the force is the 
rate at which gas transfers linear momentum to the object. But the object's 
momentum is $M V$, where we now revert to dimensional variables. In the 
reference frame where the background gas is stationary, we have
\begin{equation}
\frac{d(MV)}{dt} \,=\, -{\dot M}\,V \,\,,
\end{equation}  
which implies that
\begin{equation}
\frac{1}{V}\,\frac{dV}{dt} \,=\,-\frac{2}{M}\,\frac{dM}{dt} \,\,.
\end{equation}  
If $V_0$ and $M_0$ are the object's initial speed and mass, respectively, then
\begin{equation}
\frac{V}{V_0} \,=\,\left( \frac{M}{M_0}\right)^{-2} \,\,.
\end{equation} 

To track the speed as a function of time, we rewrite equation~(\ref{eqn:intermdot}) for the
mass accretion rate as
\begin{equation}
\frac{dM}{dt} \,=\, {{4\,\pi\,\rho_0\,c_s\,r_s^2\,
                   \left(\lambda^2\,+\,\beta^2\right)^{1/2}}\over
                   {\left(1\,+\,\beta^2\right)^2}}
                   \left({M\over M_0}\right)^2 \,\,.
\end{equation}
Here, \hbox{$\beta\,\equiv\,V/\cs$} as before, while $r_s$ is now defined in 
terms of the {\it initial} mass: \hbox{$r_s\equiv\,2\,G\,M_0/c_s^2$}. The
fully nondimensional evolutionary equation for the speed is then
\begin{equation}\label{eqn:velocity}
\left(\frac{1}{\beta}\right)\,
\frac{d\beta}{d\tau}  \,=\, -{{4\,\left(\lambda^2\,+\,\beta^2\right)^{1/2}}
\over{\left( 1\,+\,\beta^2\right)^2}} 
\left(\frac{\beta}{\beta_0}\right)^{-1/2} \,\,.
\end{equation}
In this last equation, we have introduced the initial, nondimensional speed
$\beta_0$, as well as a nondimensional time, \hbox{$\tau\,\equiv\,t/t_0$}, 
where
\begin{eqnarray}
t_0 \,&\equiv&\,\frac{c_s^3}{2\,\pi\,\rho_0\,G^2\,M_0} \\
&=\label{eqn:veltime}&\, \frac{M_0}{2\,\pi\,\rho_0\,c_s\,r_s^2} \,\,.
\end{eqnarray}
The denominator in equation (\ref{eqn:veltime}) is the fiducial mass accretion
rate defined in Section 4.1. Thus, $t_0$ is of order the accretion time onto the initial mass. 

\begin{figure}
\plotone{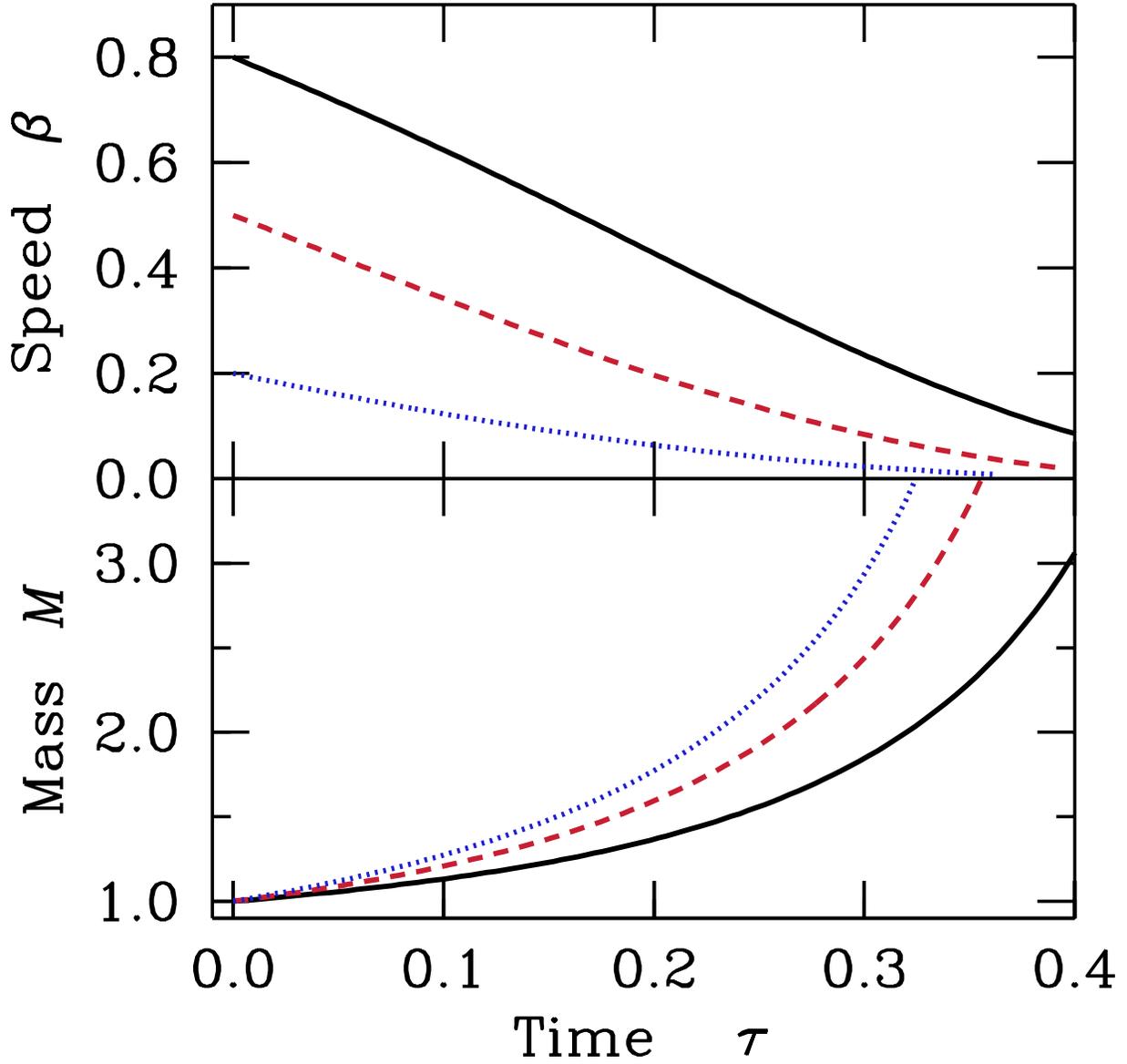}
\caption{Evolution of a particle's speed and mass as a function of nondimensional time $\tau$. The different curves represent initial speeds $\beta_0=0.8, 0.5,$ and $0.2$. A particle both triples its mass and slows to $\sim0.1$ times its initial speed in a fraction of its mass accretion time. 
}
\label{fig:massvel}
\end{figure}

The upper panel of Figure~\ref{fig:massvel} plots $\beta (\tau)$ for \hbox{$\beta_0\,=\,0.2,\,0.5$} and 0.8, 
obtained by numerical integration of equation~(\ref{eqn:velocity}). Also shown, in the lower 
panel, is the growth of the nondimensional quantity $M$, the mass of the 
gravitating body relative to its initial value. As expected, the body slows 
down appreciably within an accretion time.

\section{Summary and Discussion}
\label{sec:summary}

This study has pivoted on the close relationship between the dynamical friction
force, i.e., the transfer of linear momentum from gas to a gravitating object,
and the transfer of mass to that same object. This relationship is embodied in
our central result, equation~(\ref{eqn:mdotv}). From this equation, in turn, we derived an
analytic expression for the force itself, equation~(\ref{eqn:bondifric}).

We are now in a position to address a basic question raised in Section~2.1. How
are we justified in assuming steady-state flow, when the gravitating body is
continually decelerating? The answer is that quasi-steady flow is established
within a radius $r_{\rm crit}$ over which the sound crossing time ($r_{\rm crit}/\cs$) equals
the time for the object's momentum to decrease appreciably ($MV/F$). Recalling that $F$ is normalized to $2\pi\,\rho_0\,c^2_{\rm s}\,r^2_{\rm s}$ and using equation (\ref{eqn:mdotv}), we have, nondimensionally,
\begin{equation}
r_{\rm crit} \,=\,\alpha\,\frac{M}{\dot M} \,\,,
\end{equation}
where \hbox{$\alpha\,\equiv\, M_0/(2\,\pi\,\rho_0\,r_s^3)$}. The latter 
quantity was implicitly assumed to be large from the start, when we neglected
the self-gravity of the gas. The nondimensional mass accretion rate 
\hbox{${\dot M}\,=\,f_0 (0)$} hovers near unity for the entire evolution 
(recall~Fig.~\ref{fig:2ndg}), while $M$ itself starts at unity and climbs. Hence, the 
critical radius is much larger than $r_s$, and our analysis is 
self-consistent. 

We note that dynamical friction still operates in circumstances where mass
accretion is frustrated. For example, a wind-emitting star moving through a gas cloud experiences mass loss rather than mass gain. Cloud gas impacting the wind upstream is arrested or refracted in a bowshock, as analytically calculated by \citet{w96}. Downstream, the wind forms a supersonic jet. As long as the upstream standoff radius of the shock lies within $r_s$ and the downstream jet is relatively narrow, the far-field perturbations are close to what we have obtained, and equation~(\ref{eqn:bondifric}) for $F$ still applies. 

When the object is actually able to accept gas freely, dynamical friction 
arises in two physically distinct ways. First, there is the gravitational tug
from the wake. Second, momentum is transferred directly to the object by gas
falling onto it. Our finding that these two forces sum to ${\dot M}\,V$ is at least roughly consistent with simulations. In a numerical study directed primarily at the
mass accretion issue, \citet{r96} explicitly determined both force 
contributions on accretors of various size in a \hbox{$\gamma\,=\,1.01$} gas. For \hbox{$R/r_{\rm acc}\,=\,0.1$} and \hbox{$\beta\,=\,0.6$}, the simulation ended before the flow reached steady-state (see his Fig. 2). After initial transients died out, the gravitational drag was steady until $t\approx 13\ t_{\rm BH}$, where the Bondi-Hoyle time $t_{\rm BH}$ is $r_{\rm acc}/c_{\rm s}$. Thereafter, this force component declined for the rest of the integration. At the end of the simulation ($t = 32\ t_{\rm BH}$), the sum of the gravitational drag and momentum accretion forces was $1.2\ {\dot M}\,V$. For \hbox{$R/r_{\rm acc}\,=\,0.02$} and the same Mach number, the two forces quickly leveled off, with a sum equal to $1.4\ {\dot M}\,V$. However, this simulation ran only until $t = 10\ t_{\rm BH}$, so it is not clear whether the gravitational drag would have later declined, as in the first case.

Following historical precedent, we have restricted our investigation to an isothermal gas. For an isentropic gas with $\gamma > 1$, it seems likely that the friction force will still be given by $\dot{M}\, V$, as long as the accretor is moving subsonically. Verifying this equality analytically would require a perturbation study analogous to the present one. We leave such a project for future investigators. 

Again the current body of numerical studies is in broad accord with our expectation. \citet{r94} determined the total friction force on an accretor moving through a $\gamma=5/3$ gas. For \hbox{$R/r_{\rm acc}\,=\,0.1$} and $\beta=0.6$, the friction force was $1.1\ \dot{M}\, V$ at $t=70\ t_{\rm BH}$. For \hbox{$R/r_{\rm acc}\,=\,0.02$} and the same Mach number, the flow had not achieved steady state by $t=19\ t_{\rm BH}$. The total force was $1.8\ \dot{M}\, V$ at this time, but was falling rapidly. A future project of interest would be to redo these simulations over a range of $\beta$- and $\gamma$-values, running the simulations long enough until a true steady state is reached.

For the more general isentropic case, $\dot{M}$ can no longer be approximated by equation (\ref{eqn:intermdot}). Instead the value of $\dot{M}$ at a given $V$ decreases with higher $\gamma$-values, as shown analytically by \citet{b52} for $V=0$, and as seen in the simulations of \citet{r94,r95,r96} for accretors moving relative to the background gas. Isentropic flows are less compressible than isothermal ones, so the wake will be less dense. As a result, the friction force will also be lower, presumably by the same amount as the accretion rate $\dot{M}$. 

In the present investigation, we have been unable to tease apart analytically the two force
contributions. To do so would require study of the flow closer to the 
gravitating mass, specifically across the sonic surface. In principle, a
perturbation series in this region could be linked to the outer one developed
here. Besides elucidating the momentum transfer through infall, such a study could also establish $\dot M$ analytically as a function of velocity, thus putting accretion theory as a whole on a firmer foundation.

\acknowledgements
We gratefully acknowledge useful conversations from a number of colleagues 
during the course of the project. These include Jon Arons, Phil Chang, Chris McKee, and Prateek Sharma. We thank the referee Thiery Foglizzo for an insightful report that helped improve the clarity of our paper.  ATL acknowledges support from
an NSF Graduate Fellowship, while SWS was partially funded by 
NSF Grant~0908573.

\clearpage

\appendix

\clearpage

\end{document}